\documentclass[12pt]{article}

\RequirePackage{etex}

\usepackage[margin=1in]{geometry}  	
\usepackage{amsfonts}
\usepackage{amsmath} 
\usepackage{amssymb}
\usepackage{centernot}
\usepackage{booktabs}
\usepackage{multirow}
\usepackage{algorithm}
\usepackage{algorithmicx}
\usepackage{subcaption}
\usepackage[onehalfspacing]{setspace} 

\usepackage{float}
\usepackage{caption}
\usepackage{graphicx}
\usepackage{adjustbox}
\usepackage{lscape}

\usepackage{adjustbox}

\usepackage{xcolor}

\usepackage{relsize}

\usepackage{bbm}

\usepackage[flushleft]{threeparttable}

\definecolor{forestgreen}{RGB}{34,139,34}
\usepackage{hyperref}
\hypersetup{
    colorlinks=true,
    linkcolor=magenta,
    filecolor=magenta, 
    citecolor=forestgreen,      
    urlcolor=blue
}

\usepackage{authblk}

\usepackage{amsthm}
\newtheorem{theorem}{Theorem}

\usepackage{xpatch}
\makeatletter
\xpatchcmd{\proof}{\@addpunct{.}}{\@addpunct{:}}{}{}
\makeatother

\makeatletter
\def\@hangfrom#1{\setbox\@tempboxa\hbox{{#1}}%
      \hangindent 0pt
      \noindent\box\@tempboxa}
\makeatother

\makeatletter
\newcommand{\vast}{\bBigg@{3}}
\newcommand{\Vast}{\bBigg@{4}}
\makeatother

\makeatletter
\newcommand*{\indep}{%
  \mathbin{%
    \mathpalette{\@indep}{}%
  }%
}
\newcommand*{\nindep}{%
  \mathbin{
    \mathpalette{\@indep}{\not}
  }%
}
\newcommand*{\@indep}[2]{%
  \sbox0{$#1\perp\m@th$}
  \sbox2{$#1=$}
  \sbox4{$#1\vcenter{}$}
  \rlap{\copy0}
  \dimen@=\dimexpr\ht2-\ht4-.2pt\relax
  \kern\dimen@
  {#2}%
  \kern\dimen@
  \copy0 
} 
\makeatother

\DeclareMathOperator{\E}{\textnormal{\mbox{E}}}

\usepackage[utf8]{inputenc}
\DeclareUnicodeCharacter{200E}{}

\usepackage{tikz}
\usetikzlibrary{positioning,shapes.geometric}

\usepackage{setspace}
\usepackage{cite}

\usepackage[stable]{footmisc}

\usepackage{rotating}

\makeatletter
\def\@seccntformat#1{\@ifundefined{#1@cntformat}%
   {\csname the#1\endcsname\quad}  
   {\csname #1@cntformat\endcsname}
}
\let\oldappendix\appendix 
\renewcommand\appendix{%
    \oldappendix
    \newcommand{\section@cntformat}{\appendixname~\thesection\quad}
}
\makeatother

\usepackage[absolute,showboxes]{textpos}

\TPMargin{5pt}

\usepackage{thmtools, thm-restate}

\usepackage{datetime}

\usepackage{sectsty}
\subsectionfont{\noindent\textbf\itshape}
\subsubsectionfont{\normalfont\itshape}

\begin{document}

\title{\textbf{Robust Estimation of Loss-Based Measures of Model Performance under Covariate Shift} \vspace*{0.3in} }

\author[1]{Samantha Morrison}
\author[1]{Constantine Gatsonis}
\author[2-4]{Issa J. Dahabreh}
\author[1]{Bing Li}
\author[1]{Jon A. Steingrimsson\footnote{Address for correspondence: email: \texttt{jon\_steingrimsson@brown.edu}.}}

\affil[1]{Department of Biostatistics, School of Public Health, Brown University, Providence, RI}
\affil[2]{CAUSALab, Harvard T.H. Chan School of Public Health, Boston, MA}
\affil[3]{Department of Epidemiology, Harvard T.H. Chan School of Public Health, Boston, MA
}
\affil[4]{Department of Biostatistics, Harvard T.H. Chan School of Public Health, Boston, MA
}

\maketitle{}

\thispagestyle{empty}

\clearpage

\thispagestyle{empty}

\vspace*{1in}

\begin{abstract}  
We present methods for estimating loss-based measures of the performance of a prediction model in a target population that differs from the source population in which the model was developed, in settings where outcome and covariate data are available from the source population but only covariate data are available on a simple random sample from the target population. Prior work adjusting for differences between the two populations has used various weighting estimators with inverse odds or density ratio weights. Here, we develop more robust estimators for the target population risk (expected loss) that can be used with data-adaptive (e.g., machine learning-based) estimation of nuisance paramaters. We examine the large-sample properties of the estimators and evaluate finite sample performance in simulations. Last, we apply the methods to data from lung cancer screening using nationally representative data from the National Health and Nutrition Examination Survey (NHANES) and extend our methods to account for the complex survey design of the NHANES.
\end{abstract}

\vspace{0.5in}

\clearpage

\thispagestyle{empty}

\section{Introduction}

Ideally, a prediction model should be evaluated using data from the target population where it will be applied, but typically the source data used for model building are not obtained from a random sample of that target population \cite{cox2018big} and model performance in the source data may not reflect performance in the target population. A major reason why model performance estimates may not transport to the target population is ``covariate shift,'' that is, the presence of differences in the covariate distribution between the population underlying the source data (i.e.,~the source population) and the target population  \cite{shimodaira2000improving, bickel2009discriminative, huang2007correcting}. Under covariate shift, the conditional distribution of the outcome given the covariates is the same in both the source and target population, but the covariate distributions of the two populations are different (i.e., the populations have a different ``case-mix''). Such differences affect model performance as evaluated by measures that are averages of a loss function over the target population distribution (e.g.,~the mean squared error, MSE; mean absolute error; the Brier score \cite{brier1950verification}). 

To understand how covariate shift affects model performance, it is useful to consider ``prediction error modifiers'' -- covariates that are associated with prediction error, as assessed by a specific loss function, for a given model \cite{steingrimsson2021transporting}. When prediction error modifiers have a different distribution between the source and target population, estimators of loss-based measures of model performance that only use source data are biased for target population model performance \cite{zadrozny2004learning}. Previous efforts to correct this bias used importance-weighting approaches \cite{sugiyama2007covariate} to re-weight observations in the source data by the ratio of the covariate density of the target and source population or, equivalently, by the inverse of the conditional odds of being from the source population \cite{steingrimsson2021transporting}, in order to construct asymptotically unbiased estimators of the target population risk (i.e., the expected loss in the target population). The conditional odds (or the covariate density ratio) are almost always unknown and need to be estimated using statistical models. For the importance-weighting estimators to consistently estimate the target population risk, the models for the conditional odds needs to be correctly specified. 

Even when the model for the conditional odds of being from the source population is correctly specified, importance-weighting estimators are inefficient \cite{robins1994estimation}. Furthermore, asymptotically valid inference for importance-weighting estimators requires estimators of the conditional odds to converge at $\sqrt{n}$ rate \cite{chernozhukov2018double}. That precludes using data-adaptive estimators (e.g.,~machine learning estimators) of the density ratio or the conditional odds because such estimators converge at slower than $\sqrt{n}$ rate. Data-adaptive estimators, however, are very appealing in applied work because subject matter knowledge is typically inadequate to determine the correct specification of the conditional odds models, particularly when the covariates that differ in distribution between the source and target population are high-dimensional or have multiple continuous components.

In this paper, we develop a doubly robust estimator for the target population risk that involves estimating both the expected loss conditional on covariates and the probability of participation in the source. Our estimator is consistent for the target population risk if at least one of these models is correctly specified, but not necessarily both, and can be used for asymptotically valid inference even if the models are estimated using methods that converge at a rate slower than $\sqrt{n}$ (i.e.,~allowing the use of data-adaptive estimation). In the process of developing the doubly robust estimator, we also develop a novel conditional loss modeling-based estimator that relies on estimating the expected loss conditional on covariates. We provide identifiability conditions, identification results, and large-sample properties for the doubly robust estimator. We compare the finite-sample performance of the doubly robust and conditional loss modeling-based estimators against importance-weighting estimators in simulations. Last, we apply the methods to estimate model performance of a prediction model for lung cancer diagnosis built using data from the National Lung Screening Trial (NLST) in a target population of people eligible for lung cancer screening in the US. The sample from the target population comes from the National Health and Nutrition Examination Survey (NHANES) a complex survey that involves multi-stage clustering and variable probability sampling. We show how to modify our estimators to account for the complex sampling design and to incorporate weights that account for the oversampling of certain subgroups, survey non-response, and post-stratification adjustments.

\section{Identificaton of the target population risk \label{sec:identification}}

\subsection{Setup, study design, and targets of inference}

Let $Y$ be an outcome and $X$ a covariate vector. Let $X^*$ be a vector that contains a subset of the covariates in $X$, and $g(X^*)$ a prediction model for the conditional expectation of $Y$ given $X^*$ that we are interested in evaluating the performance of in a target population of substantive interest. We consider a setting where we have access to outcome and covariate data on a sample from the source population $\{(X_i, Y_i): i=1, \ldots, n_1\}$ and covariate data on a separately obtained random sample from the target population $\{X_i: i = 1, \ldots, n_0\}$. Let $D$ be an indicator of being from the source population (i.e.,~$D=1$ if an observation comes from the source population and $D=0$ if an observation comes from the target population). In this setup, the data used to estimate model performance in the target population is the combined source population sample and the target population sample $\{(X_i, D_i, D_i \times Y_i): i = 1, \ldots, n = n_1 + n_0\}$. We also assume that the model is fit using data that is independent of the data used to assess model performance (e.g.,~the model is an external model or it is fit using training data from the source population and an independent set of test data from the source population is used for model assessment).

We focus on etimation of loss-based measures of model performance because many popular evaluation measures are loss-based (e.g.,~mean squared error, Brier loss, and the absolute loss). A loss function $L(Y, g(X^*))$ measures the discrepancy between the observed outcome $Y$ and a prediction of the outcome $g(X^*)$. We define the target parameter of interest as the target population risk (i.e.,~expected loss in the target population):
\[
\psi = \E[L(Y,g(X^*))|D=0].
\]
Importantly, we \emph{do not} assume that the model $g(X^*)$ is correctly specified when developing our estimator, which implies that we can estimate model performance in the target population for both correctly specified and misspecified models as long as the identifiability assumptions listed below hold.

\subsection{Identification analysis}

\paragraph{Identifiability conditions:} We will argue that $\psi$ is identifiable, that is, can be written as a functional of the observed data distribution, under the following conditions which are fairly standard in the literature on transporting prediction models to a target population \cite{zadrozny2004learning,sugiyama2007covariate}:
\begin{itemize}
\label{identifability}
    \item[A1.] Conditional exchangeability over $D$: For all $x$ such that $f(X=x, D=0) \neq 0$, $$F(Y|X=x, D=0) = F(Y|X=x, D=1),$$
    where $F(Y|X=x, D=0)$ and $F(Y|X=x, D=1)$ are the cumulative distribution functions in the target and source population, respectively, and $f(X, D=0)$ is the covariate density in the target population. 
    \item[A2.] Positivity of participation in the source: For all $x$ such that $f(X=x, D=0) \neq 0$, $$\Pr[D=1|X=x] >0.$$
\end{itemize}

Assumption A1 implies that the joint distribution of $(X,Y)$ in the source and target population differ only in terms of the covariate distribution (i.e., the setting of covariate shift); however, we \emph{do not} assume that the conditional distribution of $Y$ given $X$ is known or can be consistently estimated. Assumption A2 informally says that all covariate patterns in the target population have a positive probability of appearing in the source data.

\paragraph{Identification:} We now give an identification result for the target population model performance, under conditions A1 and A2 \cite{steingrimsson2021transporting}. 

\begin{theorem}
\label{ID-Thm} If conditions A1 and A2 hold, then $\psi$ can be written as the observed data functional
\begin{equation}\label{G-ID}
    \psi = \E[L(Y, g(X^*))|D=0]  = \E[\E[L(Y, g(X^*))|X, D=1]|D=0];
\end{equation}
or, using a weighting representation,
\begin{equation}
\label{IPW-ID}
\psi = \frac{1}{\Pr[D=0]} \E\left[\frac{I(D=1) \Pr[D=0|X]}{\Pr[D=1|X]} L(Y,g(X^*)) \right].
\end{equation}
\end{theorem}

We assume that data from the source and target population are separately sampled with unknown but typically expected to be unequal sampling fractions in a ``non-nested design''.  In this situation the probabilities $\Pr[D=0|X]$ and $\Pr[D=0]$ appearing in expression \eqref{IPW-ID} are not identifiable (without additional information) \cite{dahabreh2021study, steingrimsson2021transporting} since under the non-nested design the ratio $n_0/(n_0+n_1)$ does not necessarily converge to the population probability of being in the target population, that is $\Pr[D=0]$. Nevertheless, as we discuss next, the target parameter $\psi$ is still identifiable under this biased sampling design. 

\paragraph{Identification under the biased sampling design:} Let $R$ be an indicator for sampling under the non-nested design (i.e., $R=1$ if an observation is sampled from the source or target population; otherwise $R = 0$). Because the data available are separately obtained random samples from the source population ($D=1$) and from the target population ($D=0$), we have the following theorem \cite{steingrimsson2021transporting}:
\begin{theorem}
Under the biased sampling design, the risk in the target population is identifiable by the observed data functional
\begin{equation}
\label{g-form-ID}
\psi =   \E[\E[L(Y, g(X^*))|X, D=1, R=1]|D=0,R=1].
\end{equation}
Furthermore, the right-hand-side of the above equation can be re-expressed as
\begin{equation}
\label{IPW-ID-2}
\psi = \frac{1}{\Pr[D=0|R=1]} \E\left[\frac{I(D=1) \Pr[D=0|X,R=1]}{\Pr[D=1|X,R=1]} L(Y,g(X^*)) \Bigg| R=1\right].
\end{equation}
\end{theorem}
Because all quantities in expressions \eqref{g-form-ID} and \eqref{IPW-ID-2} condition on the observed data ($R=1$), the target parameter $\psi$ is identifiable under the biased sampling design and we can interpret the expectations in Theorem \ref{ID-Thm} as being taken with respect to densities from the biased sampling design. An alternative design is a nested design where the source population is nested within a larger target population \cite{dahabreh2021study, steingrimsson2022extending}; the results presented here can be extended to nested designs.

\section{Estimation of the target population risk}

\subsection{Sample-analog estimators}

Using sample analogs in expression \eqref{G-ID} we obtain the following conditional loss-based estimator:
\begin{equation}
\label{G-est}
\widehat \psi_{CL} = \frac{1}{n_0} \sum_{i=1}^n I(D_i = 0) \widehat h(X_i),
\end{equation}
where $\widehat h(X)$ is an estimator for $\E[L(Y, g(X^*))|X, D=1]$. For simplicity of notation we suppress the dependency of $\widehat h(X)$ on $g(X^*)$ and the loss function $L(\cdot, \cdot)$.  The conditional loss estimator $\widehat \psi_{CL}$ is a consistent estimator for $\psi$ if the model $\widehat h(X)$ is correctly specified (i.e.,~$\widehat h(X)$ converges in probability to $\E[L(Y, g(X^*))|X, D=1]$).

Using sample analogs in expression \eqref{IPW-ID} we obtain the following weighting estimator \cite{robins1992recovery,horvitz1952generalization, steingrimsson2021transporting}:
\begin{equation}
\widehat \psi_{IW} = \frac{1}{n_0} \sum_{i=1}^n \frac{I(D_i = 1) (1 - \widehat p(X_i))}{\widehat p(X_i)} L(Y_i, g(X^*_i)),
\label{IPW-Est}    
\end{equation}
where $\widehat p(X)$ is an estimator for the conditional probability of belonging to the source population, $\Pr[D=1|X]$. The weighting estimator $\widehat \psi_{IW}$ is a consistent estimator for $\widehat \psi$ if the model $\widehat p(X)$ is correctly specified (i.e.,~$\widehat p(X)$ converges in probability to $\Pr[D=1|X]$). An advantage of $\widehat \psi_{IW}$ compared to the importance weighting estimator proposed by \cite{sugiyama2007covariate} is that it depends on the ratio of conditional probabilities of being in the source and the target population rather than the ratio of the source and target population densities. There are many established methods for estimating conditional probabilities with high dimensional covariates \cite{friedman2001elements}, whereas density estimation becomes challenging even with moderately dimensional covariate vectors.

\subsection{Doubly robust estimator}

The conditional loss estimator (\ref{G-est}) relies on correctly specifying the model for the conditional loss and the weighting estimator (\ref{IPW-Est}) relies on correctly specifying the model for the probability of being in the source data. Both estimators also inherit the rate of convergence of the estimator used to estimate the nuisance parameter needed for their implementation ($\Pr[D=1|X]$ or $\E[L(Y, g(X^*))|X, D=1]$). Hence if the conditional loss or weighting estimators are combined with data-adaptive estimators for $\widehat p(X)$ or $\widehat h(X)$ that converge at a rate slower than $\sqrt{n}$, then the conditional loss or weighting estimators are not $\sqrt{n}$ convergent and Wald-type confidence intervals are not asymptotically valid \cite{chernozhukov2018double}. Now we propose a doubly robust estimator that is more robust to model misspecification (i.e.,~is consistent under milder conditions) and allows for valid asymptotic inference even when data-adaptive models that are not $\sqrt{n}$ convergent are used for nuisance parameter estimation.

The doubly robust estimator combines modeling the probability of being in the source data and the model for the conditional loss. For estimating $\psi$, the doubly robust estimator is given by
\begin{align}
    \widehat{\psi}_{DR} &=  \frac{1}{n_0} \sum_{i=1}^n \Big[ I(D_i=0) \widehat{h}(X_i)    + \frac{(1 - \widehat{p}(X_i)) I(D_i=1)}{ \widehat{p}(X_i)} ( L(Y_i, g(X^*_i)) - \widehat{h}(X_i) ) \Big]. \label{eqn:dr_est} 
\end{align}
In Appendix \ref{append:deriveDR} we show how the doubly robust estimator is derived from the efficient influence function under a non-parametric model for the data \cite{fisher2020visually,van2003unified}. Of note, the estimators presented in the previous section are special cases of the doubly robust estimator: we obtain the weighting estimator by setting $\widehat{h}(X_i) = 0$ identically for all $i$; we obtain the conditional loss estimator by setting $\widehat{p}(X_i) = 1$ for all $i$.

\subsection{Asymptotic properties of the doubly robust estimator}

The presence of the estimators of nuisance parameters, $\widehat p(X)$ and $\widehat h(X)$, in the definition of $\widehat \psi_{DR}$ complicates derivation of large sample properties as standard central limit theorems and laws of large numbers are not directly applicable. To derive the large sample properties of the doubly robust estimator we define
\begin{align*}
H(p'(X), h'(X), \tau) = \tau^{-1}\left[I(D=0)h'(X) + \frac{(1-p'(X)) I(D=1)}{p'(X)}\big(L(Y, g(X^*)) - h'(X)\big)\right],
\end{align*}
for arbitrary functions $p'(X), h'(X),$ and $\tau$. The doubly robust estimator can be written as $\widehat \psi_{DR} = \frac{1}{n}\sum_{i=1}^n H(\widehat p(X_i), \widehat h(X_i), \widehat \tau)$, where $\widehat p(X_i)$, and $\widehat h(X_i)$ are as defined before, and $\widehat \tau= \frac{1}{n} \sum_{j=1}^n I(D_j=0)$. Denote the limits of $\widehat{p}(X)$ and $\widehat{h}(X)$ as $p^*(X)$ and $h^*(X)$, respectively, and let $\tau_0 = \Pr[D=0]$. Under correct model specification, the limits are equal to $p^*(X) = \Pr[D=1|X]$ and $h^*(X) = \E[L(Y,g(X^*))|X,D=1]$.

We make the following four assumptions:
\begin{itemize}
    \item[B1.]  $H(\widehat{p}(X), \widehat{h}(X), \widehat{\tau})$ and its limit $H(p^*(X), h^*(X), \tau_0)$  fall in a Donsker class \cite{dudley2014uniform}.
    \item[B2.]  $|| H(\widehat{p}(X), \widehat{h}(X), \widehat{\tau}) -  H(p^*(X), h^*(X), \tau_0)|| \overset{P}{\longrightarrow} 0$.
    \item[B3.] (Finite second moment) $\E[H(p^*(X), h^*(X), \tau_0)^2] < \infty$.
    \item[B4.] (Model double robustness) At least one of the models $\widehat p(X)$ or $\widehat h(X)$ is correctly specified. That is, at least one of $p^*(X) = \Pr[D=1|X]$ or $h^*(X) =  \E[L(Y, g(X^*))|X, D=1]$ holds, but not necessarily both.
\end{itemize}
Assumption B4 reflects the model double robustness property (i.e.,~it is enough to correctly specify one of the models $\widehat p(X)$ and $\widehat h(X)$, not both). If $\widehat p(X)$, $\widehat h(X)$, $p^*(X)$, and $h^*(X)$ are Donsker, $\widehat h(X)$ and $h^*(X)$ are uniformly bounded, and $\widehat p(X)$ and $p^*(X)$ are uniformly bounded away from zero, then Assumption B1 holds by Donsker preservation theorems \cite{kosorok2007introduction}. The Donsker requirement restricts the complexity of the models $\widehat p(X)$ and $\widehat h(X)$ and holds for many commonly used models such as generalized linear models. To relax that assumption, sample splitting can be used to allow for flexible machine learning algorithms to be used when fitting $\widehat h(X)$ and $\widehat p(X)$ without violating Assumption B1 \cite{chernozhukov2018double,bickel1988estimating,robins2008higher}. Using Assumptions B1 through B4, we prove the following theorem in the Supplementary Web Appendix:

\begin{theorem}\label{thm}
If assumptions B1 through B4 hold, then
\begin{itemize}
    \item[1.] (Consistency) $\widehat\psi_{DR} \overset{P}{\longrightarrow} \psi$.
    \item[2.] $\widehat{\psi}_{DR}$ has the asymptotic representation \begin{align*}
    \sqrt{n}(\widehat \psi_{DR} - \psi)  &=  \sqrt{n} \left(\frac{1}{n} \sum_{i=1}^n H(p^*(X_i), h^*(X_i),\tau_0) - \E\left[H(p^*(X), h^*(X), \tau_0) \right] \right) + Re + o_P(1), 
\end{align*}
where
\begin{equation}
\label{RE-term}
Re \leq \sqrt{n} \ O_P \bigg( ||\widehat{h}(X) - \E[L(Y, g(X^*))|X, D=1]||_2^2 \times ||\widehat{p}(X) -  \Pr[D=1|X]||_2^2 \bigg).
\end{equation}
\end{itemize}
\end{theorem}
Theorem \ref{thm} gives useful insights into the behaviour of the doubly robust estimator and how the models used for estimation of $\Pr[D=1|X]$ and $\E[L(Y,g(X^*))|X,D=1]$ impact the large sample properties of $\widehat \psi_{DR}$. Part 1 of Theorem \ref{thm} shows that $\widehat \psi_{DR}$ is consistent if at least one of the models $\widehat p(X)$ and $\widehat h(X)$ is correctly specified (model double robustness \cite{rotnitzky2021characterization}). In contrast, consistency of $\widehat \psi_{CL}$ relies on the estimator $\widehat h(X)$ to be correctly specified and consistency of $\widehat \psi_{IW}$ relies on the estimator $\widehat p(X)$ to be correctly specified. Furthermore, the first term on the right hand side of the asymptotic representation of $\widehat \psi_{DR}$ (in Part 2 of Theorem \ref{thm}) has the estimators $\widehat p(X)$ and $\widehat h(X)$ replaced by their limits; thus, by the Central Limit Theorem, that term converges asymptotically to a normal distribution at $\sqrt{n}$ rate.

The inequality displayed in \eqref{RE-term} shows that $\widehat \psi_{DR}$ converges at $\sqrt{n}$-rate if the combined rate of convergence of $\widehat p(X)$ and $\widehat h(X)$ is at least $\sqrt{n}$ (rate double robustness \cite{rotnitzky2021characterization}). Contrast this to $\widehat \psi_{CL}$ and $\widehat \psi_{IW}$ that inherit the rate of convergence of $\widehat h(X)$ and $\widehat p(X)$, respectively \cite{diaz2020machine}. This implies that estimation procedures such as generalized additive models and the highly adaptive Lasso \cite{benkeser2016highly} can be used to estimate $\widehat h(X)$ and $\widehat p(X)$ in $\widehat \psi_{DR}$ while still being $\sqrt{n}$ convergent (both have rate of convergence slower than $\sqrt{n}$ but faster than $n^{1/4}$ \cite{horowitz2009semiparametric,benkeser2016highly}). For inference, the variance of $\widehat \psi_{DR}$ can be estimated using either a sandwich variance estimator \cite{stefanski2002calculus} or using the bootstrap.

\section{Simulation study}
\label{sec:simulations}

\subsection{Setup}

We used simulations to compare the performance of the estimators for the target population risk $\psi$: the weighting estimator $\widehat \psi_{IW}$, the conditional loss estimator $\widehat \psi_{CL}$, the doubly robust estimator $\widehat \psi_{DR}$, and a na\"{i}ve empirical estimator that estimates the target population risk using the loss calculated using only the source data $\widehat \psi_S = \frac{1}{n_1} \sum_{i=1}^n I(D_i=1) L(Y_i, g(X^*_i))$.  

We simulated the covariate vector $X$ from a $10$ dimensional mean zero multivariate normal distribution with element $(i,j)$ of the covariance matrix equal to $0.5^{|i-j|}$. We simulated the probability of being from the source data from a Bernoulli distribution with parameter $\Pr[D=1|X] = \frac{exp(-0.3  +  0.2 \sum_{i=1}^{3} X^{(i)} + 0.3 \sum_{i=1}^3 (X^{(i)})^2)}{1 + exp(-0.3  +  0.2 \sum_{i=1}^{3} X^{(i)} + 0.3 \sum_{i=1}^3 (X^{(i)})^2)}$, where $X^{(i)}$ is the $i$th component of the covariate vector $X$. This resulted in $\Pr[D=1] = 0.61$. We simulated the outcome from a Bernoulli distribution with parameter $\Pr[Y=1|X] =\frac{exp(-0.3  +  0.2 \sum_{i=1}^{3} X^{(i)} + 0.3 \sum_{i=1}^3 (X^{(i)})^2)}{1 + exp(-0.3  +  0.2 \sum_{i=1}^{3} X^{(i)} + 0.3 \sum_{i=1}^3 (X^{(i)})^2)}$. The source data was split into a training set (2/3 of the source data) and a test set consisting of the remaining observations from the sample from the source population. The prediction model $g(X^*)$ that we assessed the performance of was a main effects logistic regression model fit using the training data from the sample from the source population. No data from the target population was used to estimate $g(X^*)$, so all observations in the target data and the test set from the sample from the source population were used to estimate the Brier loss in the target population (the evaluation measure we focus on). For each simulation we sampled a total of $1000$ observations (both $D=1$ and $D=0$), and to compare the four estimators we conducted $1000$ simulations.

For binary outcomes, expanding the square shows that to estimate $h(X) = \E[(Y-g(X^*))^2|X,D=1]$ it is enough to estimate  $\Pr[Y=1|X, D=1]$, and this is the approach we used in the simulations. To evaluate the impact of model specification on performance, we compared the four estimators using different combinations of correctly specified and misspecified models for $\Pr[Y=1|X, D=1]$ and $\Pr[D=1|X]$. For the correctly specified models, we estimated $\Pr[Y=1|X, D=1]$ and $\Pr[D=1|X]$ using a main effects logistic regression model that included both linear and quadratic main effects of all covariates. To misspecify the models for $\Pr[Y|X, D=1]$ and $\Pr[D=1|X]$ we fit a logistic regression model that only included linear main effects (i.e.,~both models failed to include the quadratic main effects). We also implemented a doubly robust estimator that used generalized additive models to estimate both $\Pr[Y=1|X, D=1]$ and  $\Pr[D=1|X]$. We fit the generalized additive models using the {\tt mgcv} package in {\tt R} \cite{wood2015package} entering all covariates as splines using the default options in the {\tt gam} function.

Based on the theoretical properties of the estimators, we expected the na\"{i}ve empirical estimator to be biased. We also expected the weighting estimator to be biased when the model for source study participation was misspecified, and the conditional loss estimator to be biased when the model for the conditional loss was incorrectly specified. Finally, we expected the doubly robust estimator to be unbiased when at least one of the models was correctly specified.

\subsection{Simulation results}
 
\begin{figure}[ht]
    \caption{Boxplots of the difference between the estimated and the true Brier loss (estimated numerically) for the na\"{i}ve empirical, weighting (IW), conditional loss (CL), and doubly robust (DR) estimators. Values closer to zero indicate better performance and the simulation setup is described in Section \ref{sec:simulations}. }
    \centering
    \includegraphics[width=\textwidth]{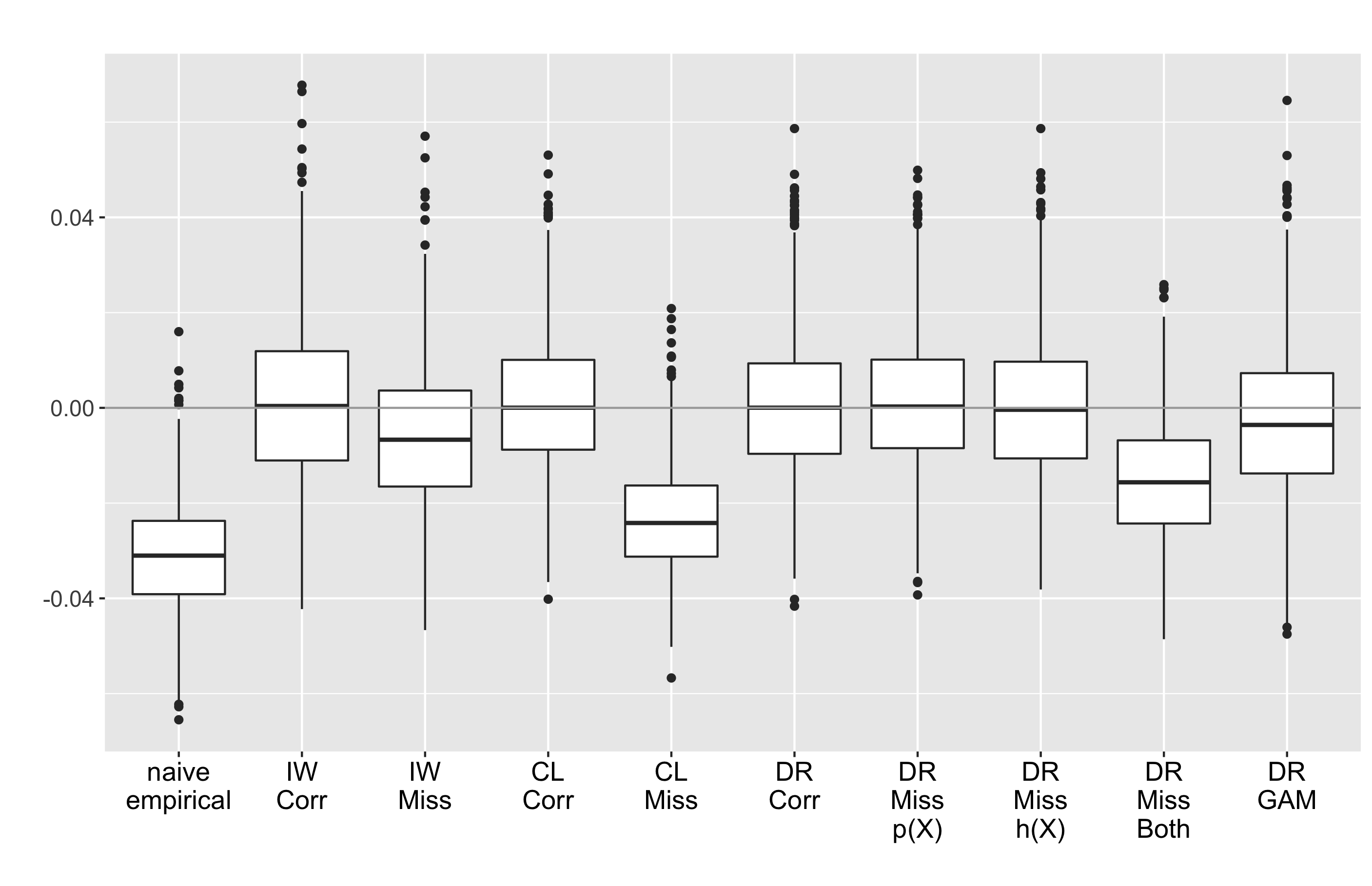}
    \caption*{Here, $h(X)$ denotes a model for $\Pr[Y=1|X, D=1]$ and $p(X)$ denotes a model for $\Pr[D=1|X]$. Corr indicates correct specification of the logistic regression models needed for implementation of the estimators. Miss indicates a misspecified logistic regression model. GAM indicates that a generalized additive model was used to estimate both $\Pr[Y=1|X, D=1]$ and  $\Pr[D=1|X]$.}
    \label{fig:classification}
\end{figure}

Figure \ref{fig:classification} shows boxplots of the results (exact numerical values are presented in Table \ref{tab:classification} in Web Appendix). As expected, the na\"{i}ve empirical estimator was biased downwards with a relative bias of $12\%$. Under misspecification of $\widehat{p}(X)$, the weighting estimator was biased but the doubly robust estimator is unbiased.  Under misspecification of $\widehat{h}(X)$, the conditional loss estimator was biased, but the doubly robust estimator was unbiased. When both models $\widehat{p}(X)$ and $\widehat{h}(X)$ were misspecified all estimators, including the doubly robust estimator, were biased. When a generalized additive model was used to estimate both $\Pr[Y=1|X, D=1]$ and  $\Pr[D=1|X]$, the doubly robust estimator was approximately unbiased (relative bias of 1.1\%). 

\section{Assessing the performance of a model for lung cancer diagnosis} \label{sec:da-l}

We applied the proposed methods to evaluate the performance of a prediction model for lung cancer diagnosis in a target population of people eligible for lung cancer screening in the US.

\subsection{Study design and data} 

We used source population data from the National Lung Screening Trial (NLST), a large clinical trial comparing the effect of screening for lung cancer using computerized tomography (CT) versus chest X-ray \cite{national2011reduced}. The NLST enrolled participants from 2002-2004 and followed them until the end of December 2009. The trial found that CT screening reduced lung cancer specific mortality and it has had a large influence on guidelines for lung cancer screening \cite{moyer2014screening, krist2021screening}. We used data from the CT arm of the NLST ($n_1 =25,825$ after removing 897 observations with missing data). We used lung cancer diagnosis within six years as the outcome. 

We defined the target population as the population of non-institutionalized U.S. adults who met the eligibility criteria of the NLST. To obtain information on the covariate distribution of the target population, we used data from the 2009-2010 cycle of NHANES, which is sampled to be representative of the non-institutionalized US population. We focused on the subset of NHANES participants that met the NLST eligibility criteria (55–74 years old with no history of lung cancer who currently smoke or have quit within the past 15 years and that have greater than 30 pack-year history) and were included in a smoking sub-study (and therefore had information on smoking habits needed to assess eligibility). This resulted in a sample size of $n_0 = 189$ after removing 4 observations with missing data. The NLST eligibility criteria are almost identical to the criteria used for recommending people for lung cancer screening in the 2014 United States Preventive Services Task Force (USPSTF) guidelines \cite{moyer2014screening} and very similar to the updated criteria in the 2021 USPSTF guidelines for lung cancer screening \cite{krist2021screening}. Because NHANES is a cross-sectional study, it does not collect long-term follow-up information on lung cancer diagnosis; that is to say, no outcome data are available in the sample from the target population. Therefore, we cannot build a prediction model and/or evaluate the performance of a prediction model using only data from NHANES.

Table \ref{eg1} in Web Appendix \ref{swa-da} summarizes variables from the NSLT and NHANES datasets that we used in our analyses. The data suggest the presence of substantial differences between the population underlying the NLST trial and the target population: participants in the NLST were less ethnically and racially diverse and are more educated compared to the national population of people eligible for NLST (represented by the subset of the NHANES data that met the NLST eligibility criteria). NHANES used a multi-cluster sampling design where each observation is associated with a sampling weight (accounting for oversampling of certain subgroups, survey non-response, and post-stratification adjustments). In subsection \ref{weigths-n} we show how the conditional loss, weighting, and doubly robust estimators can be modified to incorporate the survey sampling weights and account for the multi-stage clustering. 

\subsection{Accounting for the complex survey design of the NHANES}
\label{weigths-n}

\paragraph{Incorporating the NHANES survey sampling weights:} Let $w_i$ be the sampling weight for observation $i$ (where all observations in the sample from the source population get a weight of $1$). To account for the sampling weights, we modify the conditional loss estimator as follows:
\begin{equation}
\label{G-est-w}
\widehat \psi_{CL}^w = \frac{1}{\sum_{i=1}^n w_i I(D_i = 0)} \sum_{i=1}^n w_i I(D_i = 0) \widehat h(X_i).
\end{equation}
The estimator $\widehat h(X)$ for $\E[L(Y, g(X^*))|X, D=1]$ is obtained using only the sample from the source population; therefore, it does not need to incorporate the survey sampling weights. 

Furthermore, we modify the weighting estimator as follows:
\begin{equation}
\widehat \psi_{IW}^w =  \frac{1}{\sum_{i=1}^n w_i I(D_i = 0)} \sum_{i=1}^n \frac{I(D_i = 1) (1 - \widehat {\tilde p}(X_i))}{\widehat {\tilde p}(X_i)} L(Y_i, g(X^*_i)),
\label{IPW-Est-w}    
\end{equation}
where $\widehat {\tilde p}(X)$ is an estimator for $\Pr[D=1|X]$ that accounts for the sampling weights in the target population (e.g.,~using a weighted logistic regression model). Last, we modify the doubly robust estimator as follows:
\begin{align}
    \widehat{\psi}_{DR}^w &=  \frac{\sum_{i=1}^n \Big[ w_i I(D_i=0) \widehat{h}(X_i)    + \frac{(1 - \widehat{\tilde p}(X_i)) I(D_i=1)}{ \widehat{\tilde p}(X_i)} ( L(Y_i, g(X^*_i)) - \widehat{h}(X_i) ) \Big]}{\sum_{i=1}^n w_i I(D_i=0)} . \label{eqn:dr_est-w} 
\end{align}

\paragraph{Model specification:} We evaluated the performance a logistic regression model with the following predictors entered as main effects: age, BMI, race (Black, White, Hispanic, other), education (less than high school, high school graduate, associate's degree/some college, college graduate), personal history of cancer, smoking status, smoking intensity, duration of smoking, and smoking quit time. We split the NLST data into a training and a test set of equal size and the model was fit on the training set and the Brier score \cite{brier1950verification} in the target population was estimated using the test set from the NLST data and all the NHANES data. To implement the conditional loss, weighting, and doubly robust estimators we used a main effects logistic regression models to estimate $\Pr[D=1|X]$ weighted by the survey sampling weights; to estimate $\E[L(Y, g(X^*))|X, D=1]$ we used a main effects logistic regression model. 

\paragraph{Inference:} To quantify uncertainty about risk estimates in the target population, we used the non-parametric bootstrap with $1000$ bootstrap samples. To account for the multi-stage clustering design, we resampled the data consistent with the NHANES sampling design \cite{shao2003impact}. To obtain the standard deviation of the na\"{i}ve empirical estimator, we used the non-parametric bootstrap with $1000$ bootstrap samples.

\subsection{Results} 

Table \ref{eg2} shows estimates and associated standard errors for the Brier risk in the target population from the na\"{i}ve empirical, conditional loss, weighting, and doubly robust estimators. The conditional loss, weighting, and doubly robust estimators produced similar estimates between them, but substantially greater than those of the na\"{i}ve empirical estimator. The weighting estimator had greater standard errors than the doubly robust estimator (by 12\%) as well as the conditional loss estimator (by 53\%).

In Web Appendix \ref{ev-data} we present the results from an additional proof-of-principle analysis where the NLST data were artificially split into a sample from the source population and a sample from the target population (under the covariate shift assumption). In that analysis, by construction, outcome information was available from the target population sample, allowing us to compare the proposed estimators to an ``oracle'' estimator that used outcome information from the target population. The results show similar trends as in the simulations presented in Section \ref{sec:simulations}, the na\"{i}ve empirical estimator produced estimates that differed from those of the oracle estimator while the three estimators of the target population Brier risk that account for covariate shift produced estimates similar to the oracle estimator. Furthermore, the weighting estimator had greater variance than the doubly robust and conditional loss estimators.

\section{Discussion}

If prediction error modifiers have a different distribution between the source population from which data for model development are obtained, and the target population where the model will be applied, then measures of model performance calculated using data only from the source population may not directly apply to the target population \cite{steingrimsson2021transporting}. Hence, methods that can adjust for between-population differences in prediction error modifiers are needed when assessing model performance in the target population.

We proposed and developed theoretical properties of a doubly robust estimator for the risk in the target population that is more robust to model misspecification than the previously proposed weighting estimator and can be used with data-adaptive estimators of nuisance parameters that converge at a rate slower than $\sqrt{n}$. As a special case of the doubly robust estimator, we also proposed a novel conditional loss estimator. We compared the doubly robust and conditional loss estimators to the weighting estimator using simulations and data on lung cancer screening and found that they performed favorable compared to previously proposed inverse weighting estimators.

In our setup, the source data is split into a training set used to build the prediction model and a test set that is combined with the sample from the target population to estimate the performance of the prediction model. Alternatively, we may be interested in assessing model performance of an already established prediction model (built using an external data source). In that case, the whole sample from the source population (not just the test set) can be used for assessment of model performance and the methods developed in this manuscript are still valid.

\clearpage
\bibliographystyle{unsrt}
\bibliography{bibliography/generalizabilityBIB.bib}

\clearpage
\section*{Tables and Figures}

\vspace{0.3in}
\begin{table}[htbp]
\caption{Estimates for the Brier risk in the target population and the associated estimates of standard error estimated using data from the National Lung Screening Trial and the NHANES survey.}
\label{eg2} 
\centering 
\begin{tabular}{@{}lllll@{}}
\toprule
     &  $\widehat \psi_S$  &  $\widehat{\psi}_{IW}^w$ & $\widehat{\psi}_{CL}^w$ & $ \widehat{\psi}_{DR}^w$ \\ \midrule
Estimator for Brier risk  &  0.039 &    0.052 & 0.049 &  0.049 \\
Standard error estimator &0.0016& 0.0075 & 0.0049 & 0.0067 \\
\bottomrule
\end{tabular}
\caption*{$\widehat \psi_S$ is the na\"{i}ve empirical estimator, $\widehat{\psi}_{IW}^w$ is the weighting estimator, $\widehat{\psi}_{CL}^w$ is the conditional loss estimator, and $ \widehat{\psi}_{DR}^w$ is the doubly robust estimator.}
\end{table}

\clearpage

\vspace{0.3in}

\setcounter{section}{19}
\renewcommand{\thesection}{\Alph{section}}
\renewcommand{\thetable}{S\arabic{table}}
\renewcommand{\thefigure}{S\arabic{figure}}

\clearpage
\appendix
\setcounter{table}{0}
\setcounter{page}{1}
\section{Proofs of key results}

\subsection{Efficient influence function and doubly robust estimator \label{append:deriveDR}}

Now we will show that the influence function under a non-parametric model for the observable data is  
\begin{align*}
\chi_{p_0}^1 &=  \frac{I(D=0)}{\Pr[D=0]} \big\{E[L(Y, g(X^*))|X, D=1] - \psi \big\}  \\
 & +  \frac{\Pr[D=0|X] I(D=1)}{\Pr[D=0] \Pr[D=1|X]} \big\{ L(Y, g(X^*)) - \E[L(Y, g(X^*))|X, D=1] \big\}.
\end{align*}
As the influence function under the non-parametric model is unique it is also the efficient influence function.

To prove that $\chi_{p_0}^1$ is the influence function we show that $\chi_{p_0}^1$ is a solution to
\[
\frac{d}{dt} \psi_{p_t} \Bigr|_{t=0} = \E_{p_0}[\chi_{p_0}^1 g_{p_0}],
\]
where the left hand side of the above equation is the pathwise derivative of the target parameter and $g_{p_0}$ is the score of the observable data \cite{van2003unified}. Here, the subscript $p_0$ denotes that the expectation is calculated under the true data law. For notational convenience, define $h(X) = \E[L(Y, g(X^*))|X, D=1]$ and $O = (Y, X, D)$. 

The pathwise derivative of $\psi_{p_t}$ with respect to t where the subscript $p_t$ denotes the dependence of $\psi_{p_t}$ on a one-dimensional parametric sub-model $p_t$ indexed by $t \in [0,1]$ is given by
\begin{align*}
    \frac{\partial}{\partial t} \psi_{p_t} \Bigr|_{t=0} &= \frac{d}{dt} \E_{pt}[\E_{pt}[L(Y, g(X^*))|X, D=1]|D=0]  \Bigr|_{t=0} \\
    &= \frac{\partial}{\partial t}\E_{pt}[\E_{p_0}[L(Y, g(X^*))|X, D=1]|D=0] \Bigr|_{t=0}\\
    &+ \E_{p_0}\left[\frac{\partial}{\partial t} \E_{pt}[L(Y, g(X^*))|X, D=1]\Bigr|_{t=0} \Bigr|D=0\right]\\
    &= \E_{p_0}\left[ \big\{h(X) - \psi \big\}  \frac{I(D=0)}{\Pr[D=0]} g_{Y,X,D}(O)\right] \\
    &+ \frac{1}{\Pr[D=0]}\E_{p_0}\left[\big\{ L(Y, g(X^*)) - h(X) \big\}  \frac{I(D=1)\Pr[D=0|X]}{\Pr[D=1|X]} g_{Y, X, D}(O) \right]  \\
    &= \E_{p_0}\Bigg[\bigg( \big\{h(X) - \psi \big\} \frac{I(D=0)}{\Pr[D=0]}   \\ 
    & + \frac{1}{\Pr[D=0]} \big\{ L(Y, g(X^*)) - h(X) \big\} \Big\{ \frac{I(D=1)\Pr[D=0|X]}{\Pr[D=1|X]}\Big\}  \bigg) g_{Y, X, D}(O) \Bigg].
\end{align*}

It follows that the influence function is
\begin{align*}
\chi_{p_0}^1 &= \frac{I(D=0)}{\Pr[D=0]} \big\{E_{p_0}[L(Y, g(X^*))|X, D=1] - \psi \big\}  \\
 & +  \frac{\Pr[D=0|X] I(D=1)}{\Pr[D=0] \Pr[D=1|X]} \big\{ L(Y, g(X^*)) - \E_{p_0}[L(Y, g(X^*))|X, D=1] \big\}
\end{align*}

\subsection{Proofs of asymptotic properties of the doubly robust estimator  \label{append:asympproof}}

\paragraph{Consistency of $\widehat \psi_{DR}.$} Define the limits of $\widehat{p}(X)$ and $\widehat{h}(X)$ (assumed to exist) as $p^*(X)$ and $h^*(X)$, respectively.  Under correct model specification the limits are equal to $p^*(X) = \Pr[D=1|X]$ and $h^*(X) = \E[L(Y, g(X^*))|X, D=1]$. 

The doubly robust estimator $\widehat \psi_{DR}$ converges in probability to
\begin{align*}
\widehat{\psi}_{DR} \overset{P}{\longrightarrow} \frac{1}{\Pr[D=0]} \E\left[ I(D=0) h^*(X) + \frac{(1-p^*(X)) I(D=1)}{p^*(X)} (L(Y, g(X^*)) - h^*(X))  \right]
\end{align*}
Now we show that the quantity on the right-hand side is equal to $\psi$ under assumptions B1- B4. First consider the case where $\widehat{p}(X)$ is correctly specified, that is $p^*(X) = \Pr[D=1|X]$ (we do not make the assumption that the limit $h^*(X)$ is equal to $\E[L(Y, g(X^*))|X, D=1]$).

Algebra shows that $\psi = \frac{1}{\Pr[D=0]} \E \big[\frac{ \Pr[D=0|X] I(D=1)}{\Pr[D=1|X]} L(Y,g(X^*)) \big]$. Using that we get
   
\begin{align*}
    \widehat{\psi}_{DR} &\overset{P}{\longrightarrow} \frac{1}{\Pr[D=0]}\E \left[ I(D=0) h^*(X) + \frac{ \Pr[D=0|X] I(D=1)}{\Pr[D=1|X]} (L(Y, g(X^*)) - h^*(X))  \right] \\
    &= \frac{1}{\Pr[D=0]}\E \left[ I(D=0) h^*(X) - \frac{ \Pr[D=0|X] I(D=1)}{\Pr[D=1|X]} h^*(X)  \right]  + \psi \\
    &= \frac{1}{\Pr[D=0]}\E \left[ \E \left[ I(D=0) h^*(X) - \frac{ \Pr[D=0|X] I(D=1)}{\Pr[D=1|X]} h^*(X) \bigg|X \right] \right] + \psi \\
    &= \frac{1}{\Pr[D=0]}\E \left[ h^*(X) \E[ I(D=0)|X] - \frac{ \Pr[D=0|X]}{\Pr[D=1|X]} h^*(X) \E[I(D=1)|X] \right]  + \psi \\
    &= \frac{1}{\Pr[D=0]}\E \left[ h^*(X) \Pr[D=0|X] - \frac{ \Pr[D=0|X]}{\Pr[D=1|X]} h^*(X) \Pr[D=1|X] \right] + \psi\\
    &= \psi.
\end{align*}

Next consider the case when $\widehat{h}(X)$ is correctly specified, that is 
\[
h^*(X) = \E[L(Y, g(X^*))|X, D=1]
\]
and we do not make the assumptions that the limit $p^*(X)$ is equal to $\Pr[D=1|X]$. Algebra shows $$\psi = \frac{1}{\Pr[D=0]} \E\big[ I(D=0) \E[L(Y, g(X^*))|X, D=1] \big].$$ Hence,
\begin{align*}
    \widehat{\psi}_{DR} &\overset{P}{\longrightarrow} \frac{1}{\Pr[D=0]}\E \Bigg[ I(D=0) \E[L(Y, g(X^*))|X, D=1] \\ 
    & \quad\quad\quad + \frac{ (1-p^*(X)) I(D=1)}{p^*(X)} (L(Y, g(X^*)) - \E[L(Y, g(X^*))|X, D=1])  \Bigg] \\
    &= \psi
\end{align*}

\paragraph{Asymptotic representation of $\widehat \psi_{DR}$.}

For a random variable $W$ define 
\[\mathbb{G}_n (W) = \sqrt{n} \left(\frac{1}{n} \sum_{i=1}^n W_i - \E[W]\right).
\]
Using this notation, we rewrite the asymptotic representation as
\setcounter{equation}{0}
\renewcommand{\theequation}{S\arabic{equation}}
\begin{align}
    \sqrt{n}(\widehat{\psi}_{DR} - \psi) &= \mathbb{G}_n(H(\widehat{p}(X), \widehat{h}(X), \widehat{\tau})) - \mathbb{G}_n(H(p^*(X), h^*(X), \tau_0))  \\
    &\quad\quad\quad + \mathbb{G}_n(H(p^*(X), h^*(X), \tau_0)) \\
    &\quad\quad\quad +  \sqrt{n} \left(\E[H(\widehat{p}(X), \widehat{h}(X), \widehat{\tau})] - \psi\right) 
\end{align}

For term S1 in the above equation, the Donsker assumption (Assumption A1) implies that \cite{dudley2014uniform}
\[
\mathbb{G}_n(H(\widehat{p}(X), \widehat{h}(X), \widehat{\tau})) - \mathbb{G}_n(H(p^*(X), h^*(X), \tau_0))   = o_{P}(1).
\]
By defining
\[
Re = \sqrt{n} (\E[H(\widehat{p}(X), \widehat{h}(X), \widehat{\tau})] - \psi),
\]
we have
\[
\sqrt{n}(\widehat{\psi}_{DR} - \psi) = \sqrt{n}\left(\frac{1}{n} \sum_{i=1}^n \big( H(p^*(X_i), h^*(X_i),\tau_0) - \E\left[H(p^*(X), h^*(X), \tau_0)\right]\big) \right) + Re + o_P(1).
\]

Now we calculate the upper bound of $Re$.  First rewrite

\begin{align*}
    n^{-1/2}Re = \frac{1}{\Pr[D=0]} \bigg( \underbrace{\E\left[I(D=0)\widehat{h}(X)\right]}_{R_1} + \underbrace{\E\left[ \frac{(1-\widehat{p}(X)) I(D=1)}{\widehat{p}(X)} \{ L(Y, g(X^*)) - \widehat{h}(X) \} \right]}_{R_2} \bigg) - \psi.
\end{align*}

The first term $R_1$ can be rewritten as: 

\begin{align*}
    R_1 &= \E\left[I(D=0) \widehat{h}(X)\right] \\ &= \E\left[\E[I(D=0) \widehat{h}(X)|X]\right] \\ &=  \E\left[\E[I(D=0)|X] \widehat{h}(X)\right] \\ &= \E[\Pr[D=0|X] \widehat{h}(X)].
\end{align*}

We rewrite term $R_2$ as:
\begin{align*}
    R_2 &= \E \left[\frac{(1-\widehat{p}(X)) I(D=1)}{\widehat{p}(X)} \{ L(Y, g(X^*)) - \widehat{h}(X) \} \right] \\ &= \E \left[ \E \left[ \frac{(1-\widehat{p}(X)) I(D=1)}{\widehat{p}(X)} \{ L(Y, g(X^*)) - \widehat{h}(X) \} \bigg|X \right] \right] \ \ \ \  \\ 
    &=  \E \left[ \frac{(1-\widehat{p}(X))}{\widehat{p}(X)}  \E \left[  \frac{I(D=1)}{\Pr[D=1|X]} \Pr[D=1|X] \{ L(Y, g(X^*)) - \widehat{h}(X) \} \bigg|X \right] \right] \\
    &=  \E \left[ \frac{(1-\widehat{p}(X))}{\widehat{p}(X)} \E \left[  \Pr[D=1|X] \{ L(Y, g(X^*)) - \widehat{h}(X) \} |X, D=1 \right] \right] \\
    &=  \E \left[ \frac{(1-\widehat{p}(X))}{\widehat{p}(X)}  \Pr[D=1|X]  \{ \E \left[  L(Y, g(X^*)) |X, D=1 \right]  - \widehat{h}(X) \}  \right]
\end{align*}

Finally, we rewrite, 
\begin{align*}
    \psi &= \E[\E[L(Y, g(X^*))|X, D=1]|D=0] \\ &= \frac{1}{\Pr[D=0]} \E[I(D=0)\E[L(Y, g(X^*))|X, D=1]] \\ &= \frac{1}{\Pr[D=0]} \E[\E[I(D=0)\E[L(Y, g(X^*))|X, D=1]|X]] \\ &= \frac{1}{\Pr[D=0]} \E[\Pr[D=0|X]\E[L(Y, g(X^*))|X, D=1]] 
\end{align*}
Combining the above gives
\begin{align*}
    n^{-1/2} Re &= \frac{1}{\Pr[D=0]} \E[\Pr[D=0|X] \widehat{h}(X)] \\ &+  \frac{1}{\Pr[D=0]} \E \left[ \frac{(1-\widehat{p}(X))}{\widehat{p}(X)}  \Pr[D=1|X]  \{ \E \left[  L(Y, g(X^*)) |X, D=1 \right]  - \widehat{h}(X) \}  \right] \\ &- \frac{1}{\Pr[D=0]} \E[\Pr[D=0|X]\E[L(Y, g(X^*))|X, D=1]] \\ 
    &= \frac{1}{\Pr[D=0]} \E\big[ -\Pr[D=0|X] \{ \E[L(Y, g(X^*))|X, D=1] - \widehat{h}(X) \}  \\ &+   \frac{(1-\widehat{p}(X))}{\widehat{p}(X)}  \Pr[D=1|X]  \{ \E \big[  L(Y, g(X^*)) |X, D=1 \big]  - \widehat{h}(X) \}  \big] \\
    &= \frac{1}{\Pr[D=0]}E \Bigg[ \{ \E[L(Y, g(X^*))|X, D=1] - \widehat{h}(X) \}  \\ &\times \left\{ \frac{(1-\widehat{p}(X))}{\widehat{p}(X)}  \Pr[D=1|X]  -\Pr[D=0|X] \right\}   \Bigg] 
\end{align*}

Using the Cauchy-Schwartz inequality we get.
\begin{align*}
    Re & \le \sqrt{n} \frac{1}{\Pr[D=0]} \left(\E[\{ \E[L(Y, g(X^*))|X, D=1] - \widehat{h}(X) \}^2 ]\right)^{1/2} \\ & \times \left(\E \left[\left\{ \frac{(1-\widehat{p}(X))}{\widehat{p}(X)}  \Pr[D=1|X]  -\Pr[D=0|X] \right\}^2 \right]\right)^{1/2} \\
    \leq& \sqrt{n} \ O_P \bigg( || \E[L(Y, g(X^*))|X, D=1] - \widehat{h}(X) ||_2^2 \times ||\widehat{p}(X) -  \Pr[D=1|X]||_2^2 \bigg).
\end{align*}

If both models $\widehat p(X)$ and $\widehat h(X)$ are correctly specified and converge at a combined rate faster than $\sqrt{n}$, then $Re = o_P(1)$ and 
\begin{align*}
 \sqrt{n}(\widehat \psi_{DR} - \psi)  &=  \sqrt{n} \Bigg(\frac{1}{n} \sum_{i=1}^n H(\Pr[D=1|X_i], \E[L(Y,g(X^*))|D=1,X_i],\tau_0) \\ &- \E[H(\Pr[D=1|X], \E[L(Y,g(X^*))|D=1,X], \tau_0)] \Bigg) + o_P(1),
\end{align*}
By the central limit theorem, 
\[
\sqrt{n}\left(\frac{1}{n} \sum_{i=1}^n H(p^*(X_i), h^*(X_i),\tau_0) -  \E\left[H(p^*(X_i), h^*(X_i), \tau_0)\right] \right) \overset{d}{\longrightarrow} N(0, Var[H(p^*(X), h^*(X), \tau_0)])
\]
completing the proof.

\clearpage
\section{Additional simulation results}
\vspace{1in}
\begin{table}[ht]
\caption{Numerical results from the simulations described in Section \ref{sec:simulations} of the main text. \label{tab:classification}}
\centering
\begin{tabular}{rcccc}
  \toprule
Estimator & Average of estimates & $\sqrt{n} \times Bias$ & $\sqrt{n} \times SD$ & Relative Bias \\ 
  \midrule
Na\"{i}ve empirical & 0.2279 & -0.577 & 0.225 & -12\% \\ 
  W Corr & 0.2597 &  0.00316 & 0.285 & 0.067\% \\ 
  W Miss & 0.2520 & -0.136 & 0.264 & -2.9\% \\ 
  CL Corr & 0.2597 &  0.00379 & 0.256 & 0.080\% \\ 
  CL Miss & 0.2351 & -0.445 & 0.207 & -9.4\% \\ 
  DR Corr & 0.2595 &  0.000257 & 0.269 & 0.0054\% \\ 
 DR Miss $\widehat{p}(X)$  & 0.2596 &  0.00257 & 0.259 & 0.054\% \\ 
 DR Miss $\widehat{h}(X)$ & 0.2591 & -0.00733 & 0.268 & -0.15\% \\ 
  DR Miss Both & 0.2430 & -0.302 & 0.239 & -6.4\% \\ 
  DR GAM & 0.2565 & -0.05030 & 0.299 & -1.1\% \\ 
   \bottomrule
\end{tabular}
\caption*{Average of estimates, estimated bias, estimated standard deviation (SD), and estimated relative bias for the na\"{i}ve empirical, weighting (W), conditional loss (CL), and doubly robust (DR) estimators. The simulation setup is described in Section \ref{sec:simulations} of the main paper.  $\sqrt{n}$ is the number of observations used to fit the estimators for the target population risk (test set of $D=1$ and $D=0$).  Here, $\widehat h(X)$ is a model for $\E[L(Y,g(X^*))|X,D=1]$ and $\widehat p(X)$ denotes a model for $\Pr[D=1|X]$.  Relative bias is calculated as $\frac{estimator- truth}{truth}$. Corr indicates correct specification of all models needed for implementation of the estimator. Miss indicates a misspecified model. DR GAM indicates that a generalized additive model was used to estimate both $\E[L(Y,g(X^*))|X,D=1]$ and  $\Pr[D=1|X]$. }
\end{table}

\clearpage
\section{Additional results from transportability analyses using NLST and NHANES data}\label{swa-da}

Table \ref{eg1} shows characteristics of the variables used in the NLST to NHANES analysis presented in Section \ref{sec:da-l} in the main paper stratified by datasource.

\begin{table}[htbp] 
\small
\centering 
\caption{\small Summary of the variables used in the transportability analysis of lung cancer prediction models in Section \ref{sec:da-l} of the main paper. Continuous variables are summarized by their weighted mean (weighted standard deviation) and categorical variables are summarized by weighted percentage in each category. }
\label{eg1}
\begin{threeparttable}[htbp]
\begin{tabular}{@{}lll@{}}
\toprule
Variable & NLST & NHANES  \\ \midrule
Age & 61.4 (5.0) & 62.6(5.4)\\
BMI & 27.9 (5.1) & 29.1 (6.5)  \\
Race or ethnic group & & \\
\multicolumn{1}{l}{  Black} & 4.4\%  & 8.2\%  \\
\multicolumn{1}{l}{  White} & 89.9\%  & 81.6\% \\
\multicolumn{1}{l}{  Hispanic} & 1.8\%  & 6.2\%  \\
\multicolumn{1}{l}{  Other} & 3.8\%  &  4.0\% \\
Education level&  &\\
\multicolumn{1}{l}{  less than high school} & 6.3\%  & 36.2\%  \\
\multicolumn{1}{l}{  high school graduate} & 38.3\%  & 30.3\% \\
\multicolumn{1}{l}{  AA degree/some college} & 23.7\%  & 21.6\%  \\
\multicolumn{1}{l}{  college graduate} & 31.7\%  &  11.9\% \\
Personal history of cancer &  4.1\% & 22.3\% \\ 
Smoking status & \\
\multicolumn{1}{l}{  Current}& 48.4\% & 62.4\% \\
\multicolumn{1}{l}{  Former} &51.6\% & 37.6\%  \\
Smoking intensity \footnotemark[1] & 28.5 (11.5) & 27.2(12.1)\\
Duration of smoking (year) & 39.8 (7.3) & 42.1 (6.7) \\
Smoking quit time (year)  \footnotemark[2] & 7.3 (4.8) & 6.4 (3.7)\\

\bottomrule
\end{tabular}
 \begin{tablenotes}
 \scriptsize
     \item[1] Smoking intensity (the average number of cigarettes smoked per day) has a nonlinear association with lung cancer and was transformed by dividing by 10, exponentiating by the power -1 in the prediction model for lung cancer.
     \item[2] Smoking quit time in former smokers (years).
   \end{tablenotes}
\end{threeparttable}
\end{table}

\clearpage
\section{Evaluation of a prediction model for a positive lung screen}
\label{ev-data}

In this section, we used only data from the National Lung Screening Trial (NLST) to evaluate the performance of the different estimators for the target population risk $\psi$. Here we focused on the binary outcome of whether the baseline $T_0$ low-dose CT screen was positive (the rate of positive screens was $27.4\%$) and only used data from the CT arm of the NLST trial.



To compare the different estimators of the Brier risk in the target population, we simulated whether an observation was from the source population using two settings. In the first setting, the probability of being from the source population was $0.5$ for all covariate patterns, which implies that the covariate distribution was the same in the source population and the target population (and hence the na\"{i}ve empirical estimator is expected to be unbiased). In the second setting, the covariate distributions differed between the two populations.  The probability of being from the source population was simulated using a main effects logistic regression model with coefficients equal to $0.05$ for all covariates that had a positive association with the outcome (determined by fitting a main effects logistic regression model) and $-0.05$ for all covariates that had a negative association with the outcome (introducing systematic differences between the sample from the source population and the sample from the target population). For both settings, we simulated 1000 different splits into source and target population data.

For both settings, the prediction model, $g(X^*)$, we assessed the performance of was a main effects logistic regression model that included age at randomization, height, weight, pack years of smoking, average number of cigarettes smoked per day, age started smoking, total years of smoking, race, gender, education status, marital status, whether has smoked cigars, whether has smoked pipes, whether lives with a smoker, whether works with a smoker, whether has worked with asbestos for at least one year, whether has worked on a farm for at least one year, prior diagnosis of emphysema, prior diagnosis of hypertension, prior diagnosis of pneumonia, whether siblings have had lung cancer, and whether parents have had lung cancer. And those were also the variables we adjusted for in the transportability analysis (i.e.,~$X^* = X)$. The prediction model, $g(X^*)$, was fit using the source data training set, which was a random sample of 2/3 of the source data. Model performance was evaluated using the Brier loss estimated using the combination of the test set from the source data and all of the target data. For all the estimators, standard deviation was estimated using the non-parametric bootstrap with 1000 bootstrap samples. 

Both estimators $\widehat{p}(X)$ and $\widehat{h}(X)$ were from a linear main effects logistic regression model. As outcome information was available in the target data, we can get an unbiased estimator for $\E[(Y - g(X^*))^2|D=0]$ by using only target population data. This target population only estimator will be treated as an ``oracle'' estimator and used for comparisons to evaluate the bias (compared to the ``oracle'' estimator) of the four estimators of $\psi$. 

Table \ref{tab:NLST-results} shows results for both settings averaged across $1000$ different splits into target and source data. The estimates from the oracle estimator that used outcome data from the simulated target population was 0.197 (standard error: 0.00188) for the setting when the covariate distributions are the same in the target population and the source population and was 0.179 (SD: 0.00654) when the covariate distributions differ between the two populations.

For the setting where the covariate distribution was the same in the source and the target data, all four estimates were unbiased and had almost identical estimated standard deviation. For the setting where the covariate distributions differed, the estimates from the na\"{i}ve empirical estimator were biased (11\% relative bias) while the estimates from the other three estimators were approximately unbiased. Among the unbiased estimators, the conditional loss estimator had the smallest estimated standard deviation, followed by the doubly robust estimator, and the weighting estimator had the largest estimated standard deviation.

\clearpage
\begin{table}[ht]
\caption{Comparison of the na\"{i}ve empirical source estimator, weighting estimator, conditional loss estimator, and the doubly robust estimator using data from the National Lung Cancer Screening Trial. Bias is the difference between the average estimates from the estimator and the oracle target population estimator that used outcome information from the target population. Standard deviation is the average estimated standard deviation of the estimator (estimated using the non-parametric bootstrap with 1000 bootstrap samples). The results presented are averaged across 1000 simulated splits into a source and target data. Table (a) shows results from the simulation setting where covariate distributions in the source and target data were the same and Table (b) shows results when the two covariate distributions differed. To help with numerical comparisons, we multiplied the simulation estimates of both the bias and average standard deviation by $10^3$. 
 \vspace{0.4in}
 }
\begin{subtable}[h]{0.48\textwidth}
   \centering
\caption{Covariate distributions are the same}
\begin{adjustbox}{width = 1\textwidth}
     \begin{tabular}{rcc}
  \cline{2-3}
 & Bias & Standard deviation \\
 \cline{2-3}                               
Na\"{i}ve empirical &  0.0642 &  3.19 \\
  Weighting &  0.0514 &    3.20  \\
  Conditional loss &  0.0540 &   3.19 \\
  Doubly robust &  0.0525 &   3.20 \\ 
   \cline{2-3}
\end{tabular}
\end{adjustbox}
    \end{subtable}
    \hfill
    \begin{subtable}[h]{0.48\textwidth}
    \centering
    \caption{Covariate distributions differ} 
\begin{adjustbox}{width = 1\textwidth}
    \begin{tabular}{rll}
   \cline{2-3}
 & Bias   & Standard deviation \\
   \cline{2-3}                                              
Na\"{i}ve empirical &  18.8 & 2.29   \\
  Weighting & 0.754  & 23.2 \\
  Conditional loss & -0.498  &   57.6  \\
  Doubly robust & 1.43 &  21.2\\
  \cline{2-3}
\end{tabular}

\end{adjustbox}
\end{subtable}
\label{tab:simulated groups}
\label{tab:NLST-results}
\end{table}

\end{document}